# Hemoglobin Strain Field Waves and Allometric Functionality

V. Sachdeva and J. C. Phillips

Dept. of Physics and Astronomy, Rutgers University, Piscataway, N. J., 08854

Abstract

Hemoglobin (Hgb) forms tetramers (dimerized dimers), which enhance its globular stability and may also facilitate small gas molecule transport, as shown by recent all-atom Newtonian solvated simulations. Hydropathic bioinformatic scaling reveals many wave-like features of strained Hgb structures at the coarse-grained amino acid level, while distinguishing between these features thermodynamically. Strain fields localized near hemes interfere with extended strain fields associated with dimer interfacial misfit, resulting in wave-length dependent dimer correlation function antiresonances.

**Introduction** Hemoglobin (Hgb) transports small gas molecules ($O_2$, $CO_2$) to and from tissues, while myoglobin (Mgb) stores these molecules in tissues, and neuroglobin (Ngb) also stores them, but specifically in the central nervous system and retina. In an earlier paper [1] we showed that the evolutionary development of many subtle features of Mgb and Ngb could be monitored quantitatively by using hydropathic bioinformatic scaling. The main features of our recently developed bioinformatic scaling method were described there: amino acid (aa) specific hydropathic scales $\Psi(aa)$, and their intrachain averages over sliding windows of wave length W, $\Psi(aa,W)$.

The comparative advantage of the scaling approach is that it treats protein intrachain amino acid sequences as its primary source of highly resolved information, enabling direct connections between sequence and function, with minimal support from known secondary α helix and β strand data. It has NO adjustable parameters, and it exhibits simplified post-Newtonian dynamics. Because the scales are derived bioinformatically from the differential solvent fractal geometry of thousands of PDB structures, they very accurately incorporate universal features of allometric interactions in globular protein structures that cannot be identified easily in most Newtonian simulations. Protein network conformational changes operate in a viscoelastic conformational regime, whose reversibility is made possible by evolutionary selectivity [2]. Enhanced reversibility of the inorganic network glass transition has been reported for a few narrow alloy ranges of selected covalent glasses [3]. Of course, proteins are far more complex than glasses, and their evolutionary selectivity is the key to understanding functionality.

Allosteric aspects of globin functionality, especially the C. Bohr cooperative oxidation of tetrameric Hgb, have been the subject of 6500 papers [4]. There are many models [5,6], and depending on property and heterotropic effector, each model generates its own parameters. These are not unique, because hundreds of amino acids are involved in allosteric strain fields [6,7]. The small energies responsible for small molecule ligand-binding affinities and reaction rates were not identified with discrete structural features in early work [8].

Hgb is especially attractive because its tetrameric structure exhibits a dimerized dimer substructure, consisting of two chains, $\alpha$ (142 amino acids) and $\beta$ (147 aa). These chains dimerize as $\alpha\beta$, and the functional protein is an aggregated quaternary (dimerized dimer) $\alpha_1\beta_1\alpha_2\beta_2$, which forms a deformed globular tetrahedron. It was noted that the localized geometry of the residues forming the heme porphyrin pocket is very similar in all the globins [9]. The most conserved feature is the intrachain spacing between the distal and proximate His sites in the Hgb $\alpha,\beta$ and Mgb (154 aa) chains, which are separated by 28 aa,, while the His sites are separated by 31 aa in Ngb. This suggests a natural wave length W ~ 29 for bioinformatic scaling parameter $\Psi$(aa,W) in Hgb $\alpha,\beta$.

**Results** With bioinformatic hydrofiles we can compare Hgb $\alpha,\beta$ with Mgb and Ngb, using either the standard first-order KD hydropathicity scale [10], or the second-order fractal MZ scale [11], based on self-organized criticality [12]. Qualitatively the results (Fig. 1, with the MZ scale) are similar - Hgb $\alpha,\beta$ squeezed between Mgb and Ngb, especially in the active central region containing the distal and proximate His channel gates (sites 64 and 93 in Hgb $\beta$). As shown in Fig. 2, in this region the two Hgb $\alpha,\beta$ profiles are very similar quantitatively, including a prominent apical hydrophobic peak at 81, corresponding to a similar peak at 80 in Mgb and Ngb [1]. In Hgb$\beta$ this peak can contribute to stabilizing the globular fold, so that Lys 82 can contribute its salt bridge to binding biphosphoglycerate along with His 2 and His 143 salt bridges [13].

Is the very strong wave-sculpted Hgbα - Hgbβ similarity shown in Fig. 2 with W = 27 an accident? To answer this question, one can calculate the correlations r(W) between the two aligned Hgb α and β profiles for different values of W. As shown in Fig. 3, this correlation (averaged over the heme porphyrin pocket containing the dis and prox His) contains several surprises. For W < 27, the KD scale correlation is much larger than the MZ scale correlation, with the KD correlation peaking at 90% for W = 5. Because globins are ~ 75% α helices, and the length of the F α helix near proximate His 93 is about 10 aa, this is plausible. Hemoglobin transports small gas molecules to diverse tissues, and we saw in [1] that Mgb is slightly better described by the KD scale. Here the first-order KD scale is extremely accurate in monitoring Hgbα - Hgbβ similarity for small W.

What is most striking in Fig. 3 is the abrupt drop in r(W) above W = 27 for both KD and MZ scales, including a small peak in r(MZ) at W = 27. This says that Hgb α,β are closely similar ("coherent") for all wave lengths up to and including the heme porphyrin pocket dis - prox His separation of 27 aa. A new kind of mechanical antiresonance (dip at W = 33-35, instead of a peak) appears to be connected with oxygen cooperativity. Note that in Ngb, the dis - prox His spacing is 31. In geometrical terms, at W = 1 the correlation refers only to the central region {64,93} = {0,Z}, and its sliding window site weighting is rectangular. When W = 27, the waves are spread out, and the weighting has become triangular: it extends from –Z/2 to 3Z/2, and peaks at its center at Z/2. This suggests a mechanical interaction that involves the buffer regions {–Z/2,0} and {Z,3Z/2} connecting the heme porphyrin pocket to the Hgb α,β interfaces.

The coupling of the dimerized dimer components is known to be asymmetric, with strong interfacial $\alpha_1\beta_1$ and $\alpha_2\beta_2$ coupling, and weaker $\alpha_1\beta_2$ and $\alpha_2\beta_1$ interfacial coupling (Fig. 1 of [14]). Such asymmetric coupling is suggestive of first-order hydropathic interactions which are better described by the KD scale than by the MZ scale, as suggested by Fig. 3. For completeness the KD profiles, similar to those shown for MZ profiles in Fig. 2, are shown in Fig. 4.

**Discussion** The details of ligand migration in myoglobin and neuroglobin are relatively simple, and involve primarily two alternative paths, via either the distal or proximate His gates [1]. The situation is much more complex in the dimerized Hgb($\alpha,\beta$) dimers [15-17]. Here dimerized strains render the solvated structure more porous than the crystalline structures, and open multiple ligand tunnels, many ungated. Simulations with explicit solvent reveal specific quaternary-linked changes in $\alpha$-subunit dynamics and $\beta$-heme position, which trigger the redistribution of hot $O_2$ among its numerous successive docking sites within the tunnels [17]. These changes could occur primarily in the local heme neigHgborhood on the proximate-distal length scale, which could account for the sharp dip in Hgb$\alpha$ – Hgb$\beta$ correlation shown in Fig. 3. The porous, multiple-tunnel solvated structure is suggested to be physiologically beneficial in the crowded environment of the red blood cell [17].

The strong, narrow dip in Fig. 3 is analogous to the Fano resonances seen in nuclear, atomic and crystalline spectroscopy, which are associated with interference between localized and extended states [18]. The nonlocal connection between the porous, tunnel-ridden non-crystalline Hgb structures and the Hgb($\alpha,\beta$) correlation anitresonance in Fig. 3 suggests that cooperative hemoglobin oxidation [5] involves constructive interference of dimer hydropathic waves with W < 31, which interfere destructively at the dip. The combination of stabilizing constructive interference with destabilizing destructive interference is consistent with small molecule heme transport and dimer interfacial release. There are interesting stiffening scale-dependent systematics in nanoporous organosilicate glass thin films [19].

Softening (multiple tunnels) mixes allometric length scales. In Hgb($\alpha,\beta$) dimers the local distal-proximate length scale along the peptide chain [1] competes with the extended length scale set by the dimer interfaces. This could be the origin of the correlation antiresonance, an interference effect between localized heme strain fields and extended strain fields generated by Hgb($\alpha,\beta$) dimer interfacial misfit, which has been engineered by evolution to maximize the multiple-tunnel cooperative functionality (oxygen binding) in vivo [20]. Such interference effects may be an important factor favored by oligomerization, which here not only promotes porosity but also optimizes the multiple tunnels for ligand entrance and exit through $\alpha,\beta$ correlations.

The simplest explanation for interference effects is that intrachain hydropathic waves are real and are coupled to strain fields; in evolution-optimized proteins such interference effects need not involve quantum phases [15]. The functional significance of intrachain hydropathic waves is consistent with molecular dynamics simulations that focus on momenta rather than forces or energies [21]. The reality of hydropathic waves, caused by evolution, is supported by their weakening by synthetic mutations [22].

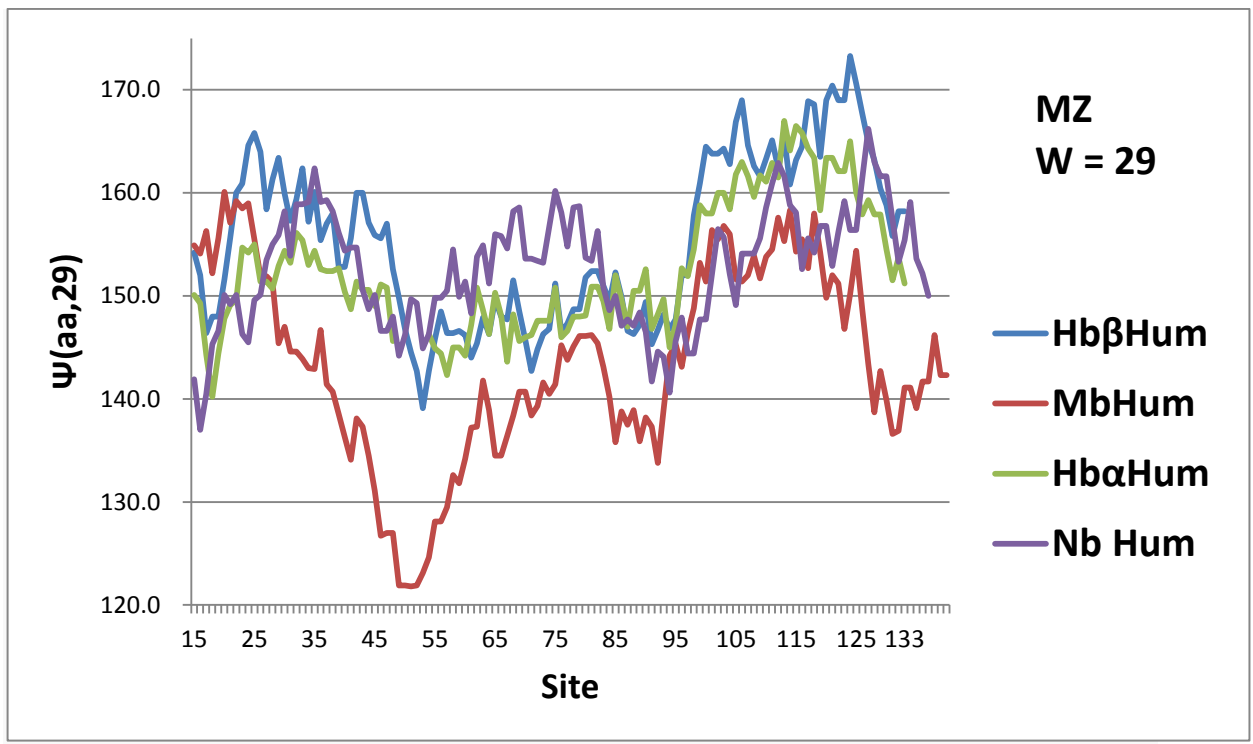

Fig. 1. A striking feature of these profiles is that Mgb is the "odd globin out", which can be understood as reflecting the function of Mgb as storing oxygen in tissues for long periods.

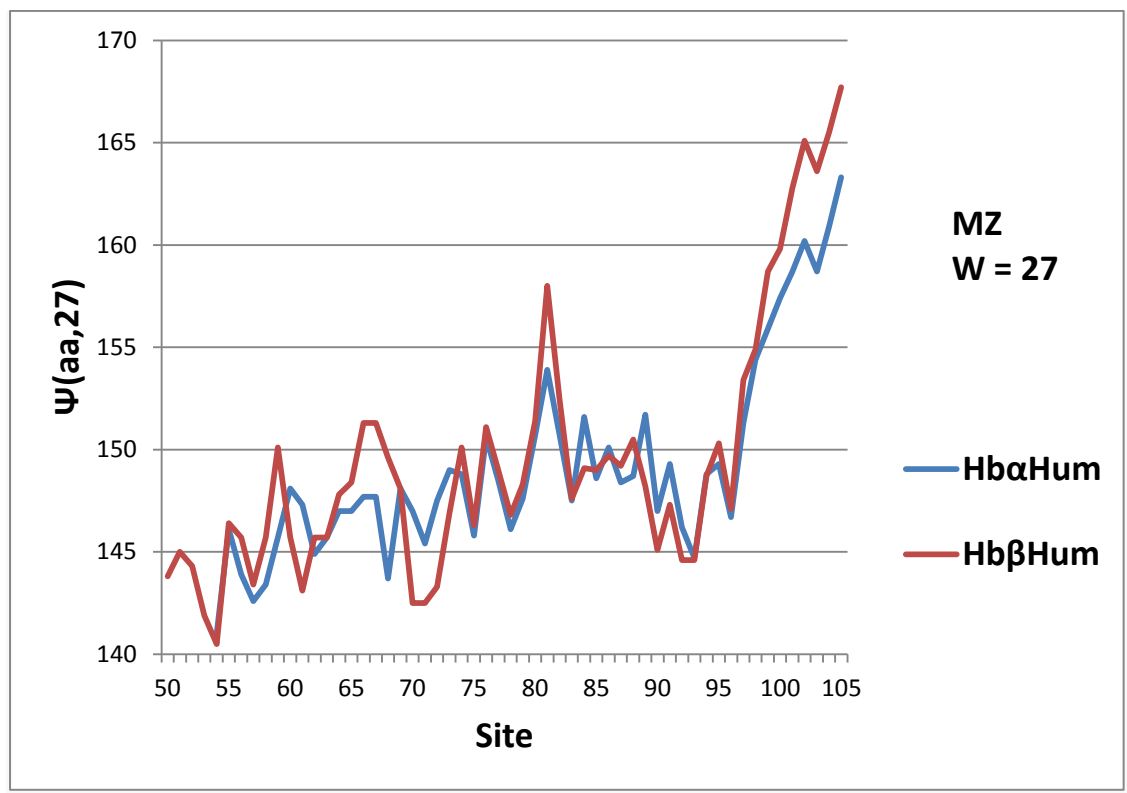

Fig. 2. There are only small overall differences between Hgbα and Hgbβ Ψ(aa,27) profiles, but it is striking that these differences are largest near the distal His 68 gate, where the Hgbβ profile has a large oscillation.  The common feature is the shared hydrophobic hinge centered between distal and proximate His gates, at 82, very similar to Mgb and Ngb (a common globin  feature).

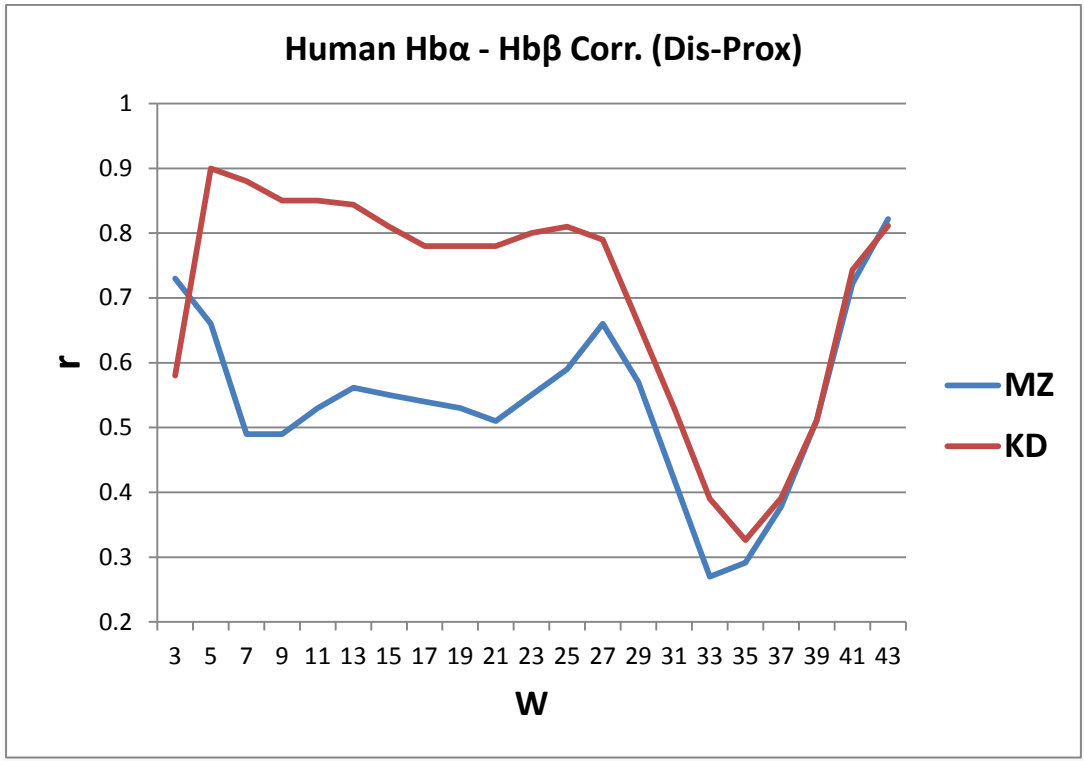

Fig. 3.  For all wavelengths W up to the His gates (Dis 68-Prox 94) spacing, Hgbα –Hgbβ are similar, especially using the KD scale.  The correlation r drops abruptly into an antiresonance for slightly longer wave lengths, and then recovers for large W.  Note that conventional BLAST comparisons correspond to W = 1 only.

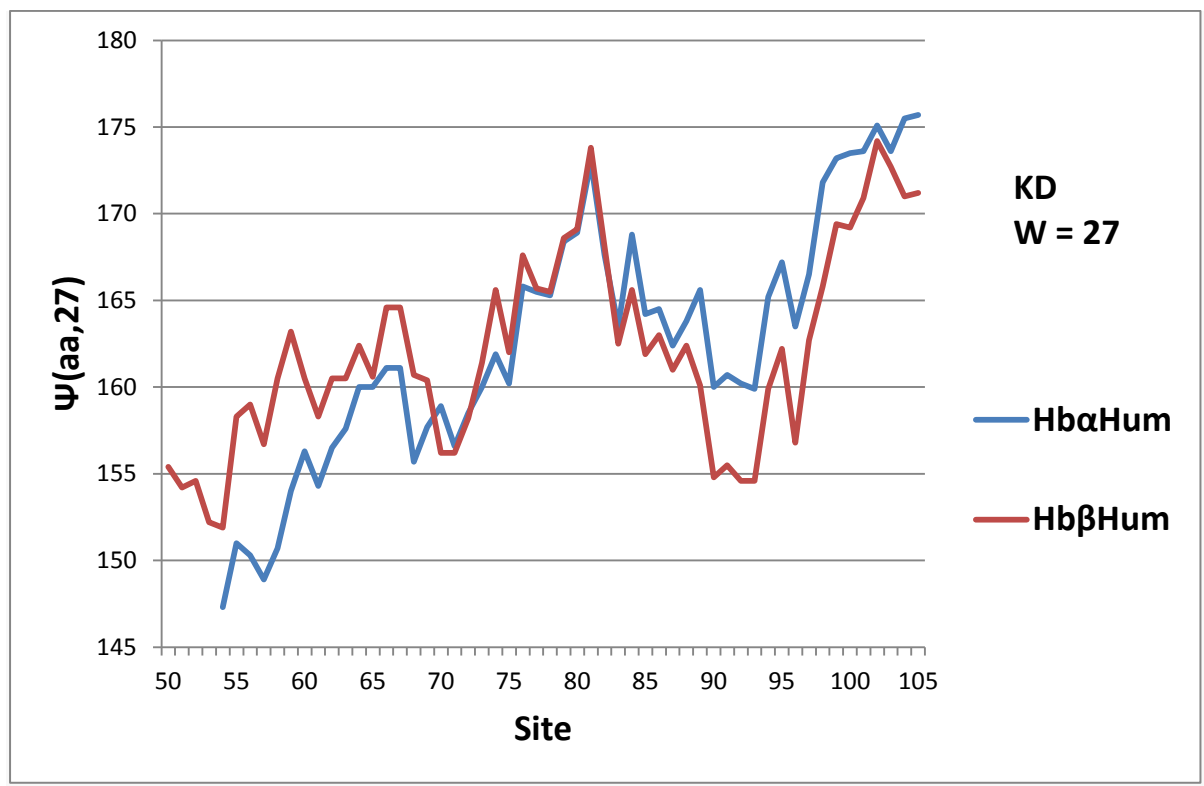

Fig. 4. This is similar to Fig. 2, except that the MZ scale there is replaced by the KD scale here. There are still only small overall differences between Hgbα and Hgbβ Ψ(aa,27) profiles, including near the distal His 68. Note the strengthening of the hydrophobic peak at 82, centered between the His gates (Dis 68-Prox 94). The hydrophilic hinge near 92 is the major difference Hgbα and Hgbβ with the KD scale.